\documentclass[twocolumn,apl,superscriptaddress]{revtex4-1}
\usepackage{amssymb}
\usepackage{amsmath}
\usepackage{graphicx}
\usepackage{wrapfig}
\usepackage{epsfig}
\usepackage{subfigure}
\usepackage[utf8]{inputenc}
\usepackage{hyperref}
\usepackage{xcolor}

\DeclareFontFamily{U}{euc}{}
\DeclareFontShape{U}{euc}{m}{n}{<-6>eurm5<6-8>eurm7<8->eurm10}{}%
\DeclareSymbolFont{AMSc}{U}{euc}{m}{n} 
\DeclareMathSymbol{\umu}{\mathord}{AMSc}{"16}

\newcommand{\rem}[1]{{\textcolor{blue}{\emph{#1}}}}

\batchmode

\begin{document}
\title{Pronounced Purcell enhancement of spontaneous emission in CdTe/ZnTe quantum dots embedded in micropillar cavities}

\author{T. Jakubczyk}
\author{W.~Pacuski}
\affiliation{%
Institute of Experimental Physics, Faculty of Physics, University of Warsaw,
Ho\.za 69, 00-681 Warsaw, Poland }%
\affiliation{%
Institute of Solid State Physics, University of Bremen,
P.O. Box 330440, 28334 Bremen, Germany}%
\author{T. Smole{\'n}ski}
\author{A. Golnik}
\affiliation{%
Institute of Experimental Physics, Faculty of Physics, University of Warsaw,
Ho\.za 69, 00-681 Warsaw, Poland }%

\author{M.~Florian}
\author{F. Jahnke}
\affiliation{%
Institute of Theoretical Physics, University of Bremen,
P.O. Box 330440, 28334 Bremen, Germany}

\author{C. Kruse}
\author{D. Hommel}
\affiliation{%
Institute of Solid State Physics, University of Bremen,
P.O. Box 330440, 28334 Bremen, Germany}%

\author{P.~Kossacki}
\affiliation{%
Institute of Experimental Physics, Faculty of Physics, University of Warsaw,
Ho\.za 69, 00-681 Warsaw, Poland }%
\date{\today}

\begin{abstract}
    The coupling of CdTe/ZnTe quantum dot (QD) emission to micropillar cavity eigenmodes in the weak coupling regime is demonstrated. We analyze photoluminescence spectra of QDs embedded in monolithic micropillar cavities based on Bragg mirrors which contain MgSe/ZnTe/MgTe superlattices as low-index material. The pillar emission shows pronounced cavity eigenmodes and their spectral shape is in good agreement with simulations. QD emission in resonance with the cavity mode is shown to be efficiently guided toward the detector and an experimental Purcell enhancement by a factor of 5.7 is determined, confirming theoretical expectations.
\end{abstract}

\maketitle

The physics of light-matter coupling in semiconductor microcavities has progressed rapidly over the last 15 years. Pillar microcavities containing QDs have been used to control spatial, energetic and temporal properties \cite{Gerard96,Gerard98} of light emission. In terms of applications, the coherent coupling between light and matter obtained in such systems is of primary importance in quantum information processing. In this context an extension of the research from the III-V based systems to II-VI based ones is promising, as robust excitonic states and a stronger carrier confinement extend the functionality range of such QD systems to higher temperatures.

Recent results on the optical control of the spin state of a single Mn atom \cite{Goryca09} underscore the perspective of enhanced light-matter coupling in ZnTe-based systems for applications in spintronics.
This material is also a good choice for devices working at the green-orange spectral region. Over the last years excellent results have been obtained for the CdTe/(CdMg)Te planar cavities \cite{Kasprzak06}, and also for the CdSe-based systems good quality micropillar cavities were demonstrated \cite{Sebald09}.
Clearly, II-VI based systems are promising candidates for the development of room temperature quantum emitters, though development of their mature cavity technology is still on a relatively early stage.

In this work we present a micro-photoluminescence ($\umu$PL) study of pillar cavities utilizing the recently developed superlattice-based distributed Bragg reflectors (DBRs) \cite{Pacuski09} lattice-matched to ZnTe \cite{Kruse11}. These DBRs contain ZnTe layers as the high refractive index material and a short-period superlattice consisting of MgSe, MgTe and ZnTe layers as the low index material. These layers have a relatively large refractive index step of $\delta n = 0.5$, which allows for the use of an efficient top DBR with a small total thickness resulting in a relatively low level of absorption of the excitation laser beam. For the resonator structure a $20$-pair bottom DBR and a $18$-pair top DBR is employed providing a high level of photon confinement within the microcavity. The cavity region sandwiched between the two DBRs is made of ZnTe and has an optical thickness of $\lambda$. A single sheet of CdTe QDs prepared by the amorphous Tellurium desorption method \cite{Tinjod03,Kruse11,Kobak11} is placed in the center of the cavity at the antinode position of the optical field. The studied micropillars of diameters ranging from $0.7$ to $2.9$ $\umu$m were etched by focused ion beam (FIB) out of the planar cavity. Lateral light confinement in the micropillar cavities is assured by the refractive index step of about $2$ between air and ZnTe~\cite{Marple64} at the investigated wavelengths. The maximum of the QDs emission intensity was designed at around $2110$~meV, while the cavity thickness, which defines the cavity mode energy, was targeted at $2050$~meV, corresponding to the low energy end of the QDs ensemble emission. This approach, after selection of an appropriate pillar, enables coupling of a single QD line with a single cavity mode.  Additionally a reference sample with nominally identical QDs in a ZnTe matrix without DBRs was grown. On the reference sample mesas of $1$~$\umu$m diameter were etched to enable spatial selection of single QDs in the spectral range of interest.

The optical measurements were done in a $\umu$PL setup. The light was focused by an aspheric lens immersed together with the sample in helium gas. The high numerical aperture of the lens (0.68) enabled collection of photons emitted at an angle of up to 42 degrees with respect to the  direction perpendicular to the sample. The lens was mounted on a positioner, which combined with an excitation spot below 2 $\umu$m enabled spatial selection of the pillar on the sample. The excitation beam was delivered from a frequency-doubled YAG laser ($532$~nm) or frequency doubled titanium-sapphire laser emitting $2$~ps pulses at a wavelength of $405$~nm. Such wavelengths are spectrally outside the DBRs stopband. The emission light was dispersed by a spectrometer and analyzed by a CCD camera or a streak camera for time-resolved signal acquisition.

\begin{figure}
	\centering
		\includegraphics[width=0.5\textwidth]{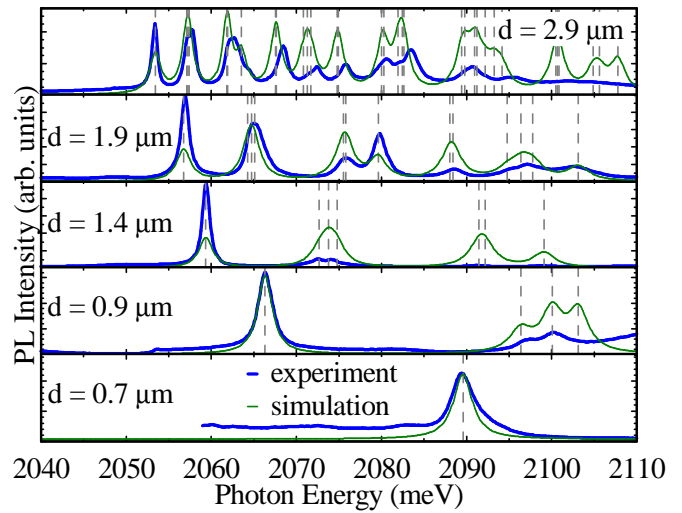}
	\caption{\rem{The experimental PL spectra of pillars with different diameters (blue lines) measured at $80$~K and under strong excitation power ($200$~$\umu$W) are compared to spectra computed using a vectorial transfer matrix approach (green lines). Please note that the theoretical spectra do not represent the PL data but rather the quantity ($1$ - Reflectivity). The vertical dashed lines indicate the energetic position of the simulated modes.}}
	\label{fig:pl_modes_v02}
\end{figure}

The $\umu$PL spectra measured at low temperature and under low excitation power (few $\umu$W) show single QD lines. At increased temperature and excitation power the broadened spectral emission of the ensemble of QDs serves as an internal light source in the investigation of cavity photonic properties \cite{Gerard96}.   Photonic eigenmodes, resulting from the optical confinement due to the DBRs and reflection from the sidewalls \cite{Jakubczyk09,Gerard96} can be observed. The typical modes are shown in Fig.~\ref{fig:pl_modes_v02} and form a characteristic pattern. With decreasing diameter of the pillar a blue-shift is observed, as well as an increase of the distance in between neighboring eigenmode lines. This is expected for increasing confinement and routinely observed for micropillar cavities in other material systems \cite{Gerard96,Sebald09}. Despite the random distribution of QDs and their emission energies the photonic pattern is reproducible for series of pillars of the same size and it undergoes smooth changes for increasing/decreasing diameter of the pillars.

In order to describe the observed cavity modes we have calculated the mode structure for the three-dimensional micropillars using a vectorial
transfer-matrix method \cite{Burak97}. This approach combines high accuracy with reasonable computational effort provided that the analysis can be restricted to the confined (bounded) modes in the transverse direction. This is well justified for pillars with sufficiently large diameters, for which the radiation in the transverse direction is negligible. Following Ref.~\cite{Burak97}, we expand the electric and magnetic fields in each layer of the
pillar with respect to cylindrical waveguide modes and determine the expansion coefficients by using the continuity relation for the field at the
interface of adjacent layers. Within the so-called common-mode approximation only waveguide modes with the same mode number are coupled across the interfaces, leading to a transfer matrix description for the light propagation through the pillar. The green lines in Fig.~\ref{fig:pl_modes_v02} show the simulated cavity spectra, wherein the refractive indices, DBR layer thicknesses, and pillar diameters enter as parameters. The refractive index of the superlattice (frequency dependent dispersion and absorption) was determined experimentally \cite{Pacuski09} and other indices were taken from Ref. \cite{Marple64}. The spectral position of the individual modes is indicated by vertical dashed lines and their broadened superposition determines the shape of the calculated cavity spectra. The obtained resonance wavelengths correspond very well to the experimental spectra, which identifies the peaks in the PL as different transverse modes of the micropillar.

\begin{figure}[h!]
	\centering
		\includegraphics[width=0.50\textwidth]{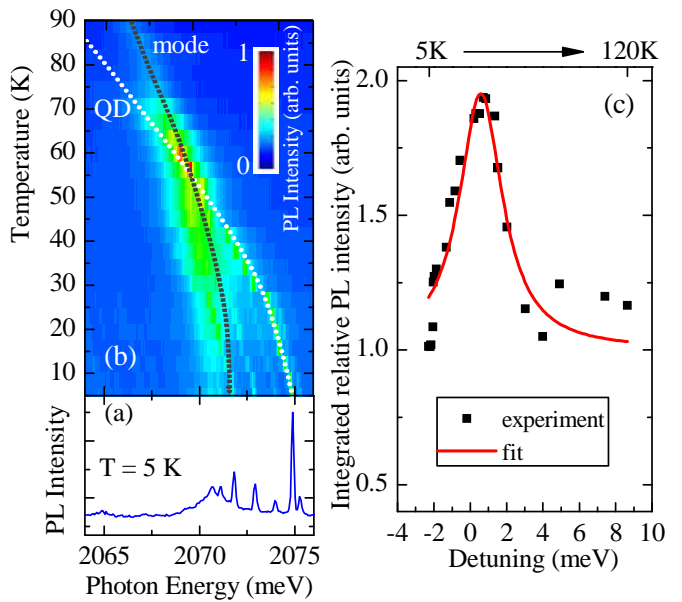}
	\caption{\rem{{\bf(a)} The cavity mode and the single QD lines are apparent in the PL spectrum of a 1.9 $\umu$m pillar for an excitation power of 50 $\umu$W at low temperature. {\bf(b)} Temperature tuning of both lines reveal resonant enhancement of the emission for zero-detuning at around 60 K. {\bf(c)} The integrated PL intensity of the QD and the cavity mode, normalized with respect to the integrated average emission of a large number of QDs emission lines and modes of the pillar, shown together with a Lorentzian fit.}}
	\label{fig:QD-mode-temp-scan}
\end{figure}
We have proven the efficient coupling of QD emission to cavity mode in the experiment with continuous wave excitation. For a pillar of 1.9 $\umu$m diameter, individual QD emission lines and eigenmode emission were identified at liquid helium temperatures (see Fig.~\ref{fig:QD-mode-temp-scan}(a)). The variation of the QD transition energy with temperature was used to control the detuning \cite{Kiraz2001}. The emission energy of QDs is following the bandgap of the material, and it depends stronger on the temperature than the energy of the photonic modes. The photoluminescence of the QD and the cavity mode as a function of their relative detuning, is shown in Fig.~\ref{fig:QD-mode-temp-scan}(b). 
From 5 K to 60 K the cavity mode shifts by about $1.5$ meV and the QDs about $5.5$ meV. At 57 K, where the energy of one of the QDs lines and the photonic cavity mode energy are equal, PL of the system shows an intensity maximum which can be regarded as a proof of resonant coupling of the QD emission to the cavity mode. In order to analyze emission intensity we integrated the intensity of the QD line and the cavity mode, and divided by the average emission of a large number of QD lines and modes of the pillar.  The obtained relative intensity (see Fig.~\ref{fig:QD-mode-temp-scan} (c)), reveals the expected Lorentzian profile as a function of temperature for both positive and negative detuning.
We pump the system well below the saturation of the QD emission. Therefore the intensity enhancement is interpreted as mostly resulting from the geometrical redirection of the emission toward the detector \cite{Munsch_PRB_09}, as pillar modes show directional emission along the growth axis \cite{Rigneault_OptLett.01,Jakubczyk11}.  The off resonant emission into the leaky modes is not collected by the detector. Additionally, we attribute the enhancement to a shortening of the lifetime of the excitonic state as a result of the Purcell effect \cite{Purcell46,Gerard98}. Faster recombination into the radiative channel of the cavity mode reduces recombination by the different ones, like other cavity modes \cite{Suffczynski09,Hennessy07} or non-radiative processes.

\begin{figure*}
	\centering
	\includegraphics{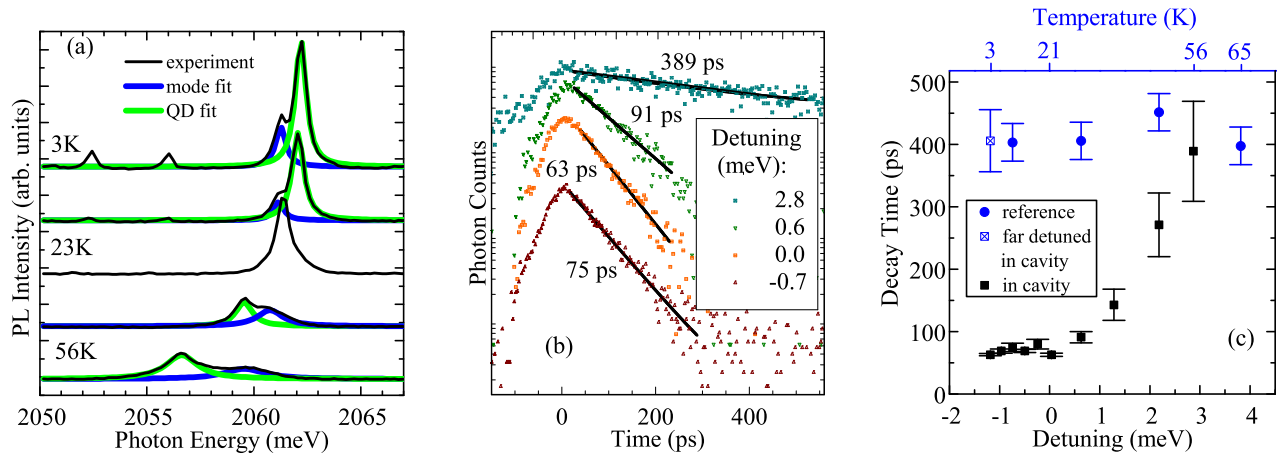}
	\caption{\rem{{\bf (a)} The PL spectra of the QD in the 1.5 $\umu$m wide pillar is shown as a function of temperature. Lorentzian fits are performed for the exciton emission line and fundamental mode emission. At low temperature two additional emission lines of far detuned dots are visible. {\bf (b)} The selected decay curves of the exciton emission in the cavity collected at various detuning are presented together with exponential fits. {\bf (c)} Decay times for the QD in cavity are shown as a function of detuning. Decay times for detuned emitters in the same cavity and reference QDs are presented as a function of temperature.}}
	\label{fig:Purcell_merged}
\end{figure*}
In order to determine the Purcell factor for our pillars, the time-resolved excitonic emission was measured as a function of the detuning from the cavity mode. We chose a single QD emission line in a micropillar of 1.4 $\umu$m diameter, which was in resonance with the fundamental mode at $23$K (see Fig.~\ref{fig:Purcell_merged}(a)). The pillar was preselected from a group of several tens of pillars as the QD showed very bright emission at resonance. This indicates a pronounced Purcell effect resulting from a good spatial matching of the QD position and the cavity mode antinode. The
choice of the resonance at low temperature limits thermally activated non-radiative recombination processes, which potentially shorten the recombination time, blurring the investigated Purcell effect. 

To explore the exciton decay dynamics we have integrated the PL in a 1 meV broad frequency window around the QD emission. As can be seen in Fig.~\ref{fig:Purcell_merged}(b) the results show an exponential decay behavior, which we have fitted accordingly and extracted the exciton lifetime. To avoid the creation of multi-exciton configurations, which would lead to a prolonged decay due to cascaded emission, we have used low excitation power.
In Fig.~\ref{fig:Purcell_merged}(c) the measured decay times are shown and exhibit a clear shortening at zero-detuning. We attribute this result to Purcell enhancement of the exciton emission. The apparent asymmetry of the decay time as a function of detuning is related to the phonon-assisted emission of the cavity mode \cite{Hohenester09}. At low temperature, where phonon emission is much more probable than phonon absorption, this emission channel is efficient only when the QD is on the higher energy side of the mode. 

For certainty that the change in the decay time is by no means related to the temperature variation we performed a similar experiment on a reference exciton line and used QD in an unstructured environment (i.e. without the cavity). A constant value of the decay time of about 400 ps was obtained. This value is in agreement with results known for CdTe/ZnTe QDs \cite{Kazimierczuk_PRB10}. Additionally we measured the average decay time of far detuned QD emission lines in the micropillar. This was done at low temperature (such lines are visible on the low energy side at Fig.~\ref{fig:Purcell_merged}(a)) and the resulting decay time is also close to 400 ps. This proves that the rate of the off-resonant emission is weakly influenced by the cavity. A pronounced photonic band gap and photonic modes are present only in the growth direction. The emitter can therefore couple to a quasi-continuum of "leaky" photon modes \cite{Bjork91,Sciesiek_APP11} and photons can be emitted in directions off the growth axis and collected due to a large numerical aperture of the lens used in the experiment (0.68).

In the case where there is no inhibition of the spontaneous emission rate for detuned emitters, the Purcell factor might be written as \cite{Gerard98}
\begin{equation}
  F = \frac{\tau (\Delta = 0)}{\tau (\Delta = \infty)}-1,
	\label{Purcell}
\end{equation}
where $\tau$ is the lifetime of the QD exciton and $\Delta$ is the detuning. In our experiment the average decay time of the reference QD is $\tau (\Delta = \infty) =(422 \pm 80)$ps and the decay time at zero detuning is $\tau (\Delta = 0) =(63 \pm 3)$ps. Thus we obtain a Purcell factor of $F =5.7 \pm 0.5$. We compare this value with a Purcell factor calculated using the formula $F=3Q\lambda_{c}^3/4\pi^{2}n^{2}V$.
We assume $Q=1900$ which is a typical quality factor observed for our micropillars with this size (for more information refer to \cite{Kruse11}). 

The mode volume $V = 28(\lambda_{c} / n)^3$ was calculated using the same extended matrix method described earlier, $n$ is the refractive index at the position of the emitter (taken from Ref. \cite{Marple64}) and $\lambda_{c}$ is the resonant wavelength of the mode. We obtain $F=5$ which is in good agreement with the experimental value. 

Concluding, we observed pronounced cavity eigenmodes as a function of the pillar diameter, showing good agreement with simulations. The emission of QDs was shown to be efficiently directionalized by means of coupling it to the cavity eigenmodes. A Purcell enhancement factor of 5.7 was determined, which is the highest reported, experimentally obtained value for II-VI based micropillars, confirmed by calculations.
Clearly, ZnTe - based monolithic micropillars efficiently modify the emission of CdTe QDs showing their potential as efficient, high-temperature single photon emitters. After further enhancement of the cavity - QD coupling various cavity quantum electrodynamics experiments are expected to be feasible on this recently developed system.
	
The authors gratefully acknowledge financial support by the DAAD which supported TJ's stay in Bremen, by the National Science Center in Poland (DEC-2011/01/N/ST3/04536 and DEC-2011/02/A/ST3/00131), by the BMBF via QK QuaHL-Rep (01BQ1037) and a "Lider" grant from The National Center for Research and Development in Poland.


\begin{thebibliography}{23}%
\makeatletter
\providecommand \@ifxundefined [1]{%
 \@ifx{#1\undefined}
}%
\providecommand \@ifnum [1]{%
 \ifnum #1\expandafter \@firstoftwo
 \else \expandafter \@secondoftwo
 \fi
}%
\providecommand \@ifx [1]{%
 \ifx #1\expandafter \@firstoftwo
 \else \expandafter \@secondoftwo
 \fi
}%
\providecommand \natexlab [1]{#1}%
\providecommand \enquote  [1]{``#1''}%
\providecommand \bibnamefont  [1]{#1}%
\providecommand \bibfnamefont [1]{#1}%
\providecommand \citenamefont [1]{#1}%
\providecommand \href@noop [0]{\@secondoftwo}%
\providecommand \href [0]{\begingroup \@sanitize@url \@href}%
\providecommand \@href[1]{\@@startlink{#1}\@@href}%
\providecommand \@@href[1]{\endgroup#1\@@endlink}%
\providecommand \@sanitize@url [0]{\catcode `\\12\catcode `\$12\catcode
  `\&12\catcode `\#12\catcode `\^12\catcode `\_12\catcode `\%12\relax}%
\providecommand \@@startlink[1]{}%
\providecommand \@@endlink[0]{}%
\providecommand \url  [0]{\begingroup\@sanitize@url \@url }%
\providecommand \@url [1]{\endgroup\@href {#1}{\urlprefix }}%
\providecommand \urlprefix  [0]{URL }%
\providecommand \Eprint [0]{\href }%
\providecommand \doibase [0]{http://dx.doi.org/}%
\providecommand \selectlanguage [0]{\@gobble}%
\providecommand \bibinfo  [0]{\@secondoftwo}%
\providecommand \bibfield  [0]{\@secondoftwo}%
\providecommand \translation [1]{[#1]}%
\providecommand \BibitemOpen [0]{}%
\providecommand \bibitemStop [0]{}%
\providecommand \bibitemNoStop [0]{.\EOS\space}%
\providecommand \EOS [0]{\spacefactor3000\relax}%
\providecommand \BibitemShut  [1]{\csname bibitem#1\endcsname}%
\let\auto@bib@innerbib\@empty
\bibitem [{\citenamefont {G\'{e}rard}\ \emph {et~al.}(1996)\citenamefont
  {G\'{e}rard}, \citenamefont {Barrier}, \citenamefont {Marzin}, \citenamefont
  {Kuszelewicz}, \citenamefont {Manin}, \citenamefont {Costard}, \citenamefont
  {Thierry-Mieg},\ and\ \citenamefont {Rivera}}]{Gerard96}%
  \BibitemOpen
  \bibfield  {author} {\bibinfo {author} {\bibfnamefont {J.~M.}\ \bibnamefont
  {G\'{e}rard}}, \bibinfo {author} {\bibfnamefont {D.}~\bibnamefont {Barrier}},
  \bibinfo {author} {\bibfnamefont {J.~Y.}\ \bibnamefont {Marzin}}, \bibinfo
  {author} {\bibfnamefont {R.}~\bibnamefont {Kuszelewicz}}, \bibinfo {author}
  {\bibfnamefont {L.}~\bibnamefont {Manin}}, \bibinfo {author} {\bibfnamefont
  {E.}~\bibnamefont {Costard}}, \bibinfo {author} {\bibfnamefont
  {V.}~\bibnamefont {Thierry-Mieg}}, \ and\ \bibinfo {author} {\bibfnamefont
  {T.}~\bibnamefont {Rivera}},\ }\href {\doibase 10.1063/1.118135} {\bibfield
  {journal} {\bibinfo  {journal} {Applied Physics Letters}\ }\textbf {\bibinfo
  {volume} {69}},\ \bibinfo {pages} {449} (\bibinfo {year} {1996})}\BibitemShut
  {NoStop}%
\bibitem [{\citenamefont {G\'erard}\ \emph {et~al.}(1998)\citenamefont
  {G\'erard}, \citenamefont {Sermage}, \citenamefont {Gayral}, \citenamefont
  {Legrand}, \citenamefont {Costard},\ and\ \citenamefont
  {Thierry-Mieg}}]{Gerard98}%
  \BibitemOpen
  \bibfield  {author} {\bibinfo {author} {\bibfnamefont {J.~M.}\ \bibnamefont
  {G\'erard}}, \bibinfo {author} {\bibfnamefont {B.}~\bibnamefont {Sermage}},
  \bibinfo {author} {\bibfnamefont {B.}~\bibnamefont {Gayral}}, \bibinfo
  {author} {\bibfnamefont {B.}~\bibnamefont {Legrand}}, \bibinfo {author}
  {\bibfnamefont {E.}~\bibnamefont {Costard}}, \ and\ \bibinfo {author}
  {\bibfnamefont {V.}~\bibnamefont {Thierry-Mieg}},\ }\href {\doibase
  10.1103/PhysRevLett.81.1110} {\bibfield  {journal} {\bibinfo  {journal}
  {Phys. Rev. Lett.}\ }\textbf {\bibinfo {volume} {81}},\ \bibinfo {pages}
  {1110} (\bibinfo {year} {1998})}\BibitemShut {NoStop}%
\bibitem [{\citenamefont {Goryca}\ \emph {et~al.}(2009)\citenamefont {Goryca},
  \citenamefont {Kazimierczuk}, \citenamefont {Nawrocki}, \citenamefont
  {Golnik}, \citenamefont {Kossacki}, \citenamefont {Wojnar}, \citenamefont
  {Karczewski},\ and\ \citenamefont {Gaj}}]{Goryca09}%
  \BibitemOpen
  \bibfield  {author} {\bibinfo {author} {\bibfnamefont {M.}~\bibnamefont
  {Goryca}}, \bibinfo {author} {\bibfnamefont {T.}~\bibnamefont
  {Kazimierczuk}}, \bibinfo {author} {\bibfnamefont {M.}~\bibnamefont
  {Nawrocki}}, \bibinfo {author} {\bibfnamefont {A.}~\bibnamefont {Golnik}},
  \bibinfo {author} {\bibfnamefont {P.}~\bibnamefont {Kossacki}}, \bibinfo
  {author} {\bibfnamefont {P.}~\bibnamefont {Wojnar}}, \bibinfo {author}
  {\bibfnamefont {G.}~\bibnamefont {Karczewski}}, \ and\ \bibinfo {author}
  {\bibfnamefont {J.~A.}\ \bibnamefont {Gaj}},\ }\href {\doibase
  10.1103/PhysRevLett.103.087401} {\bibfield  {journal} {\bibinfo  {journal}
  {Phys. Rev. Lett.}\ }\textbf {\bibinfo {volume} {103}},\ \bibinfo {pages}
  {087401} (\bibinfo {year} {2009})}\BibitemShut {NoStop}%
\bibitem [{\citenamefont {Kasprzak}\ \emph {et~al.}(2006)\citenamefont
  {Kasprzak}, \citenamefont {Richard}, \citenamefont {Kundermann},
  \citenamefont {Baas}, \citenamefont {Jeambrun}, \citenamefont {Keeling},
  \citenamefont {Marchetti}, \citenamefont {Szymanska}, \citenamefont {Andre},
  \citenamefont {Staehli}, \citenamefont {Savona}, \citenamefont {Littlewood},
  \citenamefont {Deveaud},\ and\ \citenamefont {Dang}}]{Kasprzak06}%
  \BibitemOpen
  \bibfield  {author} {\bibinfo {author} {\bibfnamefont {J.}~\bibnamefont
  {Kasprzak}}, \bibinfo {author} {\bibfnamefont {M.}~\bibnamefont {Richard}},
  \bibinfo {author} {\bibfnamefont {S.}~\bibnamefont {Kundermann}}, \bibinfo
  {author} {\bibfnamefont {A.}~\bibnamefont {Baas}}, \bibinfo {author}
  {\bibfnamefont {P.}~\bibnamefont {Jeambrun}}, \bibinfo {author}
  {\bibfnamefont {J.~M.~J.}\ \bibnamefont {Keeling}}, \bibinfo {author}
  {\bibfnamefont {F.~M.}\ \bibnamefont {Marchetti}}, \bibinfo {author}
  {\bibfnamefont {M.~H.}\ \bibnamefont {Szymanska}}, \bibinfo {author}
  {\bibfnamefont {R.}~\bibnamefont {Andre}}, \bibinfo {author} {\bibfnamefont
  {J.~L.}\ \bibnamefont {Staehli}}, \bibinfo {author} {\bibfnamefont
  {V.}~\bibnamefont {Savona}}, \bibinfo {author} {\bibfnamefont {P.~B.}\
  \bibnamefont {Littlewood}}, \bibinfo {author} {\bibfnamefont
  {B.}~\bibnamefont {Deveaud}}, \ and\ \bibinfo {author} {\bibfnamefont
  {L.~S.}\ \bibnamefont {Dang}},\ }\href {\doibase 10.1038/nature05131}
  {\bibfield  {journal} {\bibinfo  {journal} {Nature}\ }\textbf {\bibinfo
  {volume} {443}},\ \bibinfo {pages} {409} (\bibinfo {year}
  {2006})}\BibitemShut {NoStop}%
\bibitem [{\citenamefont {Sebald}\ \emph {et~al.}(2009)\citenamefont {Sebald},
  \citenamefont {Kruse},\ and\ \citenamefont {Wiersig}}]{Sebald09}%
  \BibitemOpen
  \bibfield  {author} {\bibinfo {author} {\bibfnamefont {K.}~\bibnamefont
  {Sebald}}, \bibinfo {author} {\bibfnamefont {C.}~\bibnamefont {Kruse}}, \
  and\ \bibinfo {author} {\bibfnamefont {J.}~\bibnamefont {Wiersig}},\ }\href
  {\doibase 10.1002/pssb.200844194} {\bibfield  {journal} {\bibinfo  {journal}
  {physica status solidi (b)}\ }\textbf {\bibinfo {volume} {246}},\ \bibinfo
  {pages} {255} (\bibinfo {year} {2009})}\BibitemShut {NoStop}%
\bibitem [{\citenamefont {Pacuski}\ \emph {et~al.}(2009)\citenamefont
  {Pacuski}, \citenamefont {Kruse}, \citenamefont {Figge},\ and\ \citenamefont
  {Hommel}}]{Pacuski09}%
  \BibitemOpen
  \bibfield  {author} {\bibinfo {author} {\bibfnamefont {W.}~\bibnamefont
  {Pacuski}}, \bibinfo {author} {\bibfnamefont {C.}~\bibnamefont {Kruse}},
  \bibinfo {author} {\bibfnamefont {S.}~\bibnamefont {Figge}}, \ and\ \bibinfo
  {author} {\bibfnamefont {D.}~\bibnamefont {Hommel}},\ }\href {\doibase
  10.1063/1.3136755} {\bibfield  {journal} {\bibinfo  {journal} {Applied
  Physics Letters}\ }\textbf {\bibinfo {volume} {94}},\ \bibinfo {eid} {191108}
  (\bibinfo {year} {2009})}\BibitemShut {NoStop}%
\bibitem [{\citenamefont {Kruse}\ \emph {et~al.}(2011)\citenamefont {Kruse},
  \citenamefont {Pacuski}, \citenamefont {Jakubczyk}, \citenamefont {Kobak},
  \citenamefont {Gaj}, \citenamefont {Frank}, \citenamefont {Schowalter},
  \citenamefont {Rosenauer}, \citenamefont {Florian}, \citenamefont {Jahnke},\
  and\ \citenamefont {Hommel}}]{Kruse11}%
  \BibitemOpen
  \bibfield  {author} {\bibinfo {author} {\bibfnamefont {C.}~\bibnamefont
  {Kruse}}, \bibinfo {author} {\bibfnamefont {W.}~\bibnamefont {Pacuski}},
  \bibinfo {author} {\bibfnamefont {T.}~\bibnamefont {Jakubczyk}}, \bibinfo
  {author} {\bibfnamefont {J.}~\bibnamefont {Kobak}}, \bibinfo {author}
  {\bibfnamefont {J.~A.}\ \bibnamefont {Gaj}}, \bibinfo {author} {\bibfnamefont
  {K.}~\bibnamefont {Frank}}, \bibinfo {author} {\bibfnamefont
  {M.}~\bibnamefont {Schowalter}}, \bibinfo {author} {\bibfnamefont
  {A.}~\bibnamefont {Rosenauer}}, \bibinfo {author} {\bibfnamefont
  {M.}~\bibnamefont {Florian}}, \bibinfo {author} {\bibfnamefont
  {F.}~\bibnamefont {Jahnke}}, \ and\ \bibinfo {author} {\bibfnamefont
  {D.}~\bibnamefont {Hommel}},\ }\href
  {http://stacks.iop.org/0957-4484/22/i=28/a=285204} {\bibfield  {journal}
  {\bibinfo  {journal} {Nanotechnology}\ }\textbf {\bibinfo {volume} {22}},\
  \bibinfo {pages} {285204} (\bibinfo {year} {2011})}\BibitemShut {NoStop}%
\bibitem [{\citenamefont {Tinjod}\ \emph {et~al.}(2003)\citenamefont {Tinjod},
  \citenamefont {Gilles}, \citenamefont {Moehl}, \citenamefont {Kheng},\ and\
  \citenamefont {Mariette}}]{Tinjod03}%
  \BibitemOpen
  \bibfield  {author} {\bibinfo {author} {\bibfnamefont {F.}~\bibnamefont
  {Tinjod}}, \bibinfo {author} {\bibfnamefont {B.}~\bibnamefont {Gilles}},
  \bibinfo {author} {\bibfnamefont {S.}~\bibnamefont {Moehl}}, \bibinfo
  {author} {\bibfnamefont {K.}~\bibnamefont {Kheng}}, \ and\ \bibinfo {author}
  {\bibfnamefont {H.}~\bibnamefont {Mariette}},\ }\href {\doibase
  10.1063/1.1583141} {\bibfield  {journal} {\bibinfo  {journal} {Appl. Phys.
  Lett.}\ }\textbf {\bibinfo {volume} {82}},\ \bibinfo {pages} {4340} (\bibinfo
  {year} {2003})}\BibitemShut {NoStop}%
\bibitem [{\citenamefont {Kobak}\ \emph {et~al.}(2011)\citenamefont {Kobak},
  \citenamefont {Pacuski}, \citenamefont {Jakubczyk}, \citenamefont
  {Kazimierczuk}, \citenamefont {Golnik}, \citenamefont {Frank}, \citenamefont
  {Rosenauer}, \citenamefont {Kruse}, \citenamefont {Hommel},\ and\
  \citenamefont {Gaj}}]{Kobak11}%
  \BibitemOpen
  \bibfield  {author} {\bibinfo {author} {\bibfnamefont {J.}~\bibnamefont
  {Kobak}}, \bibinfo {author} {\bibfnamefont {W.}~\bibnamefont {Pacuski}},
  \bibinfo {author} {\bibfnamefont {T.}~\bibnamefont {Jakubczyk}}, \bibinfo
  {author} {\bibfnamefont {T.}~\bibnamefont {Kazimierczuk}}, \bibinfo {author}
  {\bibfnamefont {A.}~\bibnamefont {Golnik}}, \bibinfo {author} {\bibfnamefont
  {K.}~\bibnamefont {Frank}}, \bibinfo {author} {\bibfnamefont
  {A.}~\bibnamefont {Rosenauer}}, \bibinfo {author} {\bibfnamefont
  {C.}~\bibnamefont {Kruse}}, \bibinfo {author} {\bibfnamefont
  {D.}~\bibnamefont {Hommel}}, \ and\ \bibinfo {author} {\bibfnamefont {J.~A.}\
  \bibnamefont {Gaj}},\ }\href@noop {} {\bibfield  {journal} {\bibinfo
  {journal} {Acta Physica Polonica A}\ }\textbf {\bibinfo {volume} {119}},\
  \bibinfo {pages} {627} (\bibinfo {year} {2011})}\BibitemShut {NoStop}%
\bibitem [{\citenamefont {Marple}(1964)}]{Marple64}%
  \BibitemOpen
  \bibfield  {author} {\bibinfo {author} {\bibfnamefont {D.~T.~F.}\
  \bibnamefont {Marple}},\ }\href {\doibase 10.1063/1.1713411} {\bibfield
  {journal} {\bibinfo  {journal} {J. Appl. Phys.}\ }\textbf {\bibinfo {volume}
  {35}},\ \bibinfo {pages} {539} (\bibinfo {year} {1964})}\BibitemShut
  {NoStop}%
\bibitem [{\citenamefont {Jakubczyk}\ \emph {et~al.}(2009)\citenamefont
  {Jakubczyk}, \citenamefont {Kazimierczuk}, \citenamefont {Golnik},
  \citenamefont {Bienias}, \citenamefont {Pacuski}, \citenamefont {Kruse},
  \citenamefont {Hommel}, \citenamefont {Klopotowski}, \citenamefont
  {Wojtowicz},\ and\ \citenamefont {Gaj}}]{Jakubczyk09}%
  \BibitemOpen
  \bibfield  {author} {\bibinfo {author} {\bibfnamefont {T.}~\bibnamefont
  {Jakubczyk}}, \bibinfo {author} {\bibfnamefont {T.}~\bibnamefont
  {Kazimierczuk}}, \bibinfo {author} {\bibfnamefont {A.}~\bibnamefont
  {Golnik}}, \bibinfo {author} {\bibfnamefont {P.}~\bibnamefont {Bienias}},
  \bibinfo {author} {\bibfnamefont {W.}~\bibnamefont {Pacuski}}, \bibinfo
  {author} {\bibfnamefont {C.}~\bibnamefont {Kruse}}, \bibinfo {author}
  {\bibfnamefont {D.}~\bibnamefont {Hommel}}, \bibinfo {author} {\bibfnamefont
  {L.}~\bibnamefont {Klopotowski}}, \bibinfo {author} {\bibfnamefont
  {T.}~\bibnamefont {Wojtowicz}}, \ and\ \bibinfo {author} {\bibfnamefont
  {J.~A.}\ \bibnamefont {Gaj}},\ }\href@noop {} {\bibfield  {journal} {\bibinfo
   {journal} {Acta Physica Polonica}\ }\textbf {\bibinfo {volume} {116}},\
  \bibinfo {pages} {888} (\bibinfo {year} {2009})}\BibitemShut {NoStop}%
\bibitem [{\citenamefont {{Burak}}\ and\ \citenamefont
  {{Binder}}(1997)}]{Burak97}%
  \BibitemOpen
  \bibfield  {author} {\bibinfo {author} {\bibfnamefont {D.}~\bibnamefont
  {{Burak}}}\ and\ \bibinfo {author} {\bibfnamefont {R.}~\bibnamefont
  {{Binder}}},\ }\href {\doibase 10.1109/3.594886} {\bibfield  {journal}
  {\bibinfo  {journal} {IEEE Journal of Quantum Electronics}\ }\textbf
  {\bibinfo {volume} {33}},\ \bibinfo {pages} {1205} (\bibinfo {year}
  {1997})}\BibitemShut {NoStop}%
\bibitem [{\citenamefont {Kiraz}\ \emph {et~al.}(2001)\citenamefont {Kiraz},
  \citenamefont {Michler}, \citenamefont {Becher}, \citenamefont {Gayral},
  \citenamefont {Imamoglu}, \citenamefont {Zhang}, \citenamefont {Hu},
  \citenamefont {Schoenfeld},\ and\ \citenamefont {Petroff}}]{Kiraz2001}%
  \BibitemOpen
  \bibfield  {author} {\bibinfo {author} {\bibfnamefont {A.}~\bibnamefont
  {Kiraz}}, \bibinfo {author} {\bibfnamefont {P.}~\bibnamefont {Michler}},
  \bibinfo {author} {\bibfnamefont {C.}~\bibnamefont {Becher}}, \bibinfo
  {author} {\bibfnamefont {B.}~\bibnamefont {Gayral}}, \bibinfo {author}
  {\bibfnamefont {A.}~\bibnamefont {Imamoglu}}, \bibinfo {author}
  {\bibfnamefont {L.}~\bibnamefont {Zhang}}, \bibinfo {author} {\bibfnamefont
  {E.}~\bibnamefont {Hu}}, \bibinfo {author} {\bibfnamefont {W.~V.}\
  \bibnamefont {Schoenfeld}}, \ and\ \bibinfo {author} {\bibfnamefont {P.~M.}\
  \bibnamefont {Petroff}},\ }\href {\doibase 10.1063/1.1379987} {\bibfield
  {journal} {\bibinfo  {journal} {Applied Physics Letters}\ }\textbf {\bibinfo
  {volume} {78}},\ \bibinfo {pages} {3932} (\bibinfo {year}
  {2001})}\BibitemShut {NoStop}%
\bibitem [{\citenamefont {Munsch}\ \emph {et~al.}(2009)\citenamefont {Munsch},
  \citenamefont {Mosset}, \citenamefont {Auff\`eves}, \citenamefont {Seidelin},
  \citenamefont {Poizat}, \citenamefont {G\'erard}, \citenamefont {Lemaitre},
  \citenamefont {Sagnes},\ and\ \citenamefont {Senellart}}]{Munsch_PRB_09}%
  \BibitemOpen
  \bibfield  {author} {\bibinfo {author} {\bibfnamefont {M.}~\bibnamefont
  {Munsch}}, \bibinfo {author} {\bibfnamefont {A.}~\bibnamefont {Mosset}},
  \bibinfo {author} {\bibfnamefont {A.}~\bibnamefont {Auff\`eves}}, \bibinfo
  {author} {\bibfnamefont {S.}~\bibnamefont {Seidelin}}, \bibinfo {author}
  {\bibfnamefont {J.~P.}\ \bibnamefont {Poizat}}, \bibinfo {author}
  {\bibfnamefont {J.-M.}\ \bibnamefont {G\'erard}}, \bibinfo {author}
  {\bibfnamefont {A.}~\bibnamefont {Lemaitre}}, \bibinfo {author}
  {\bibfnamefont {I.}~\bibnamefont {Sagnes}}, \ and\ \bibinfo {author}
  {\bibfnamefont {P.}~\bibnamefont {Senellart}},\ }\href {\doibase
  10.1103/PhysRevB.80.115312} {\bibfield  {journal} {\bibinfo  {journal} {Phys.
  Rev. B}\ }\textbf {\bibinfo {volume} {80}},\ \bibinfo {pages} {115312}
  (\bibinfo {year} {2009})}\BibitemShut {NoStop}%
\bibitem [{\citenamefont {Rigneault}\ \emph {et~al.}(2001)\citenamefont
  {Rigneault}, \citenamefont {Broudic}, \citenamefont {Gayral},\ and\
  \citenamefont {G\'{e}rard}}]{Rigneault_OptLett.01}%
  \BibitemOpen
  \bibfield  {author} {\bibinfo {author} {\bibfnamefont {H.}~\bibnamefont
  {Rigneault}}, \bibinfo {author} {\bibfnamefont {J.}~\bibnamefont {Broudic}},
  \bibinfo {author} {\bibfnamefont {B.}~\bibnamefont {Gayral}}, \ and\ \bibinfo
  {author} {\bibfnamefont {J.~M.}\ \bibnamefont {G\'{e}rard}},\ }\href
  {\doibase 10.1364/OL.26.001595} {\bibfield  {journal} {\bibinfo  {journal}
  {Opt. Lett.}\ }\textbf {\bibinfo {volume} {26}},\ \bibinfo {pages} {1595}
  (\bibinfo {year} {2001})}\BibitemShut {NoStop}%
\bibitem [{\citenamefont {Jakubczyk}\ \emph {et~al.}(2011)\citenamefont
  {Jakubczyk}, \citenamefont {Pacuski}, \citenamefont {Duch}, \citenamefont
  {Godlewski}, \citenamefont {Golnik}, \citenamefont {Kruse}, \citenamefont
  {Hommel},\ and\ \citenamefont {Gaj}}]{Jakubczyk11}%
  \BibitemOpen
  \bibfield  {author} {\bibinfo {author} {\bibfnamefont {T.}~\bibnamefont
  {Jakubczyk}}, \bibinfo {author} {\bibfnamefont {W.}~\bibnamefont {Pacuski}},
  \bibinfo {author} {\bibfnamefont {P.}~\bibnamefont {Duch}}, \bibinfo {author}
  {\bibfnamefont {P.}~\bibnamefont {Godlewski}}, \bibinfo {author}
  {\bibfnamefont {A.}~\bibnamefont {Golnik}}, \bibinfo {author} {\bibfnamefont
  {C.}~\bibnamefont {Kruse}}, \bibinfo {author} {\bibfnamefont
  {D.}~\bibnamefont {Hommel}}, \ and\ \bibinfo {author} {\bibfnamefont
  {J.}~\bibnamefont {Gaj}},\ }\href
  {http://dx.doi.org/10.2478/s11534-010-0131-8} {\bibfield  {journal} {\bibinfo
   {journal} {Central European Journal of Physics}\ }\textbf {\bibinfo {volume}
  {9}},\ \bibinfo {pages} {428} (\bibinfo {year} {2011})},\ \bibinfo {note}
  {10.2478/s11534-010-0131-8}\BibitemShut {NoStop}%
\bibitem [{\citenamefont {{Purcell}}(1946)}]{Purcell46}%
  \BibitemOpen
  \bibfield  {author} {\bibinfo {author} {\bibfnamefont {E.~M.}\ \bibnamefont
  {{Purcell}}},\ }\href@noop {} {\bibfield  {journal} {\bibinfo  {journal}
  {Physical Review}\ }\textbf {\bibinfo {volume} {69}},\ \bibinfo {pages} {681}
  (\bibinfo {year} {1946})}\BibitemShut {NoStop}%
\bibitem [{\citenamefont {Suffczy\ifmmode~\acute{n}\else \'{n}\fi{}ski}\ \emph
  {et~al.}(2009)\citenamefont {Suffczy\ifmmode~\acute{n}\else \'{n}\fi{}ski},
  \citenamefont {Dousse}, \citenamefont {Gauthron}, \citenamefont
  {Lema\^\i{}tre}, \citenamefont {Sagnes}, \citenamefont {Lanco}, \citenamefont
  {Bloch}, \citenamefont {Voisin},\ and\ \citenamefont
  {Senellart}}]{Suffczynski09}%
  \BibitemOpen
  \bibfield  {author} {\bibinfo {author} {\bibfnamefont {J.}~\bibnamefont
  {Suffczy\ifmmode~\acute{n}\else \'{n}\fi{}ski}}, \bibinfo {author}
  {\bibfnamefont {A.}~\bibnamefont {Dousse}}, \bibinfo {author} {\bibfnamefont
  {K.}~\bibnamefont {Gauthron}}, \bibinfo {author} {\bibfnamefont
  {A.}~\bibnamefont {Lema\^\i{}tre}}, \bibinfo {author} {\bibfnamefont
  {I.}~\bibnamefont {Sagnes}}, \bibinfo {author} {\bibfnamefont
  {L.}~\bibnamefont {Lanco}}, \bibinfo {author} {\bibfnamefont
  {J.}~\bibnamefont {Bloch}}, \bibinfo {author} {\bibfnamefont
  {P.}~\bibnamefont {Voisin}}, \ and\ \bibinfo {author} {\bibfnamefont
  {P.}~\bibnamefont {Senellart}},\ }\href {\doibase
  10.1103/PhysRevLett.103.027401} {\bibfield  {journal} {\bibinfo  {journal}
  {Phys. Rev. Lett.}\ }\textbf {\bibinfo {volume} {103}},\ \bibinfo {pages}
  {027401} (\bibinfo {year} {2009})}\BibitemShut {NoStop}%
\bibitem [{\citenamefont {Hennessy}\ \emph {et~al.}(2007)\citenamefont
  {Hennessy}, \citenamefont {Badolato}, \citenamefont {Winger}, \citenamefont
  {Gerace}, \citenamefont {Atature}, \citenamefont {Gulde}, \citenamefont
  {Falt}, \citenamefont {Hu},\ and\ \citenamefont {Imamoglu}}]{Hennessy07}%
  \BibitemOpen
  \bibfield  {author} {\bibinfo {author} {\bibfnamefont {K.}~\bibnamefont
  {Hennessy}}, \bibinfo {author} {\bibfnamefont {A.}~\bibnamefont {Badolato}},
  \bibinfo {author} {\bibfnamefont {M.}~\bibnamefont {Winger}}, \bibinfo
  {author} {\bibfnamefont {D.}~\bibnamefont {Gerace}}, \bibinfo {author}
  {\bibfnamefont {M.}~\bibnamefont {Atature}}, \bibinfo {author} {\bibfnamefont
  {S.}~\bibnamefont {Gulde}}, \bibinfo {author} {\bibfnamefont
  {S.}~\bibnamefont {Falt}}, \bibinfo {author} {\bibfnamefont {E.~L.}\
  \bibnamefont {Hu}}, \ and\ \bibinfo {author} {\bibfnamefont {A.}~\bibnamefont
  {Imamoglu}},\ }\href {\doibase 10.1038/nature05586} {\bibfield  {journal}
  {\bibinfo  {journal} {Nature}\ }\textbf {\bibinfo {volume} {445}},\ \bibinfo
  {pages} {896} (\bibinfo {year} {2007})}\BibitemShut {NoStop}%
\bibitem [{\citenamefont {Hohenester}\ \emph {et~al.}(2009)\citenamefont
  {Hohenester}, \citenamefont {Laucht}, \citenamefont {Kaniber}, \citenamefont
  {Hauke}, \citenamefont {Neumann}, \citenamefont {Mohtashami}, \citenamefont
  {Seliger}, \citenamefont {Bichler},\ and\ \citenamefont
  {Finley}}]{Hohenester09}%
  \BibitemOpen
  \bibfield  {author} {\bibinfo {author} {\bibfnamefont {U.}~\bibnamefont
  {Hohenester}}, \bibinfo {author} {\bibfnamefont {A.}~\bibnamefont {Laucht}},
  \bibinfo {author} {\bibfnamefont {M.}~\bibnamefont {Kaniber}}, \bibinfo
  {author} {\bibfnamefont {N.}~\bibnamefont {Hauke}}, \bibinfo {author}
  {\bibfnamefont {A.}~\bibnamefont {Neumann}}, \bibinfo {author} {\bibfnamefont
  {A.}~\bibnamefont {Mohtashami}}, \bibinfo {author} {\bibfnamefont
  {M.}~\bibnamefont {Seliger}}, \bibinfo {author} {\bibfnamefont
  {M.}~\bibnamefont {Bichler}}, \ and\ \bibinfo {author} {\bibfnamefont
  {J.~J.}\ \bibnamefont {Finley}},\ }\href {\doibase
  10.1103/PhysRevB.80.201311} {\bibfield  {journal} {\bibinfo  {journal} {Phys.
  Rev. B}\ }\textbf {\bibinfo {volume} {80}},\ \bibinfo {pages} {201311}
  (\bibinfo {year} {2009})}\BibitemShut {NoStop}%
\bibitem [{\citenamefont {Kazimierczuk}\ \emph {et~al.}(2010)\citenamefont
  {Kazimierczuk}, \citenamefont {Goryca}, \citenamefont {Koperski},
  \citenamefont {Golnik}, \citenamefont {Gaj}, \citenamefont {Nawrocki},
  \citenamefont {Wojnar},\ and\ \citenamefont {Kossacki}}]{Kazimierczuk_PRB10}%
  \BibitemOpen
  \bibfield  {author} {\bibinfo {author} {\bibfnamefont {T.}~\bibnamefont
  {Kazimierczuk}}, \bibinfo {author} {\bibfnamefont {M.}~\bibnamefont
  {Goryca}}, \bibinfo {author} {\bibfnamefont {M.}~\bibnamefont {Koperski}},
  \bibinfo {author} {\bibfnamefont {A.}~\bibnamefont {Golnik}}, \bibinfo
  {author} {\bibfnamefont {J.~A.}\ \bibnamefont {Gaj}}, \bibinfo {author}
  {\bibfnamefont {M.}~\bibnamefont {Nawrocki}}, \bibinfo {author}
  {\bibfnamefont {P.}~\bibnamefont {Wojnar}}, \ and\ \bibinfo {author}
  {\bibfnamefont {P.}~\bibnamefont {Kossacki}},\ }\href {\doibase
  10.1103/PhysRevB.81.155313} {\bibfield  {journal} {\bibinfo  {journal} {Phys.
  Rev. B}\ }\textbf {\bibinfo {volume} {81}},\ \bibinfo {pages} {155313}
  (\bibinfo {year} {2010})}\BibitemShut {NoStop}%
\bibitem [{\citenamefont {Bj\"ork}\ \emph {et~al.}(1991)\citenamefont
  {Bj\"ork}, \citenamefont {Machida}, \citenamefont {Yamamoto},\ and\
  \citenamefont {Igeta}}]{Bjork91}%
  \BibitemOpen
  \bibfield  {author} {\bibinfo {author} {\bibfnamefont {G.}~\bibnamefont
  {Bj\"ork}}, \bibinfo {author} {\bibfnamefont {S.}~\bibnamefont {Machida}},
  \bibinfo {author} {\bibfnamefont {Y.}~\bibnamefont {Yamamoto}}, \ and\
  \bibinfo {author} {\bibfnamefont {K.}~\bibnamefont {Igeta}},\ }\href
  {\doibase 10.1103/PhysRevA.44.669} {\bibfield  {journal} {\bibinfo  {journal}
  {Phys. Rev. A}\ }\textbf {\bibinfo {volume} {44}},\ \bibinfo {pages} {669}
  (\bibinfo {year} {1991})}\BibitemShut {NoStop}%
\bibitem [{\citenamefont {\'Sciesiek}\ \emph {et~al.}(2011)\citenamefont
  {\'Sciesiek}, \citenamefont {Gietka}, \citenamefont {Golnik}, \citenamefont
  {Kossacki}, \citenamefont {Jakubczyk}, \citenamefont {Pacuski}, \citenamefont
  {Kruse},\ and\ \citenamefont {Hommel}}]{Sciesiek_APP11}%
  \BibitemOpen
  \bibfield  {author} {\bibinfo {author} {\bibfnamefont {M.}~\bibnamefont
  {\'Sciesiek}}, \bibinfo {author} {\bibfnamefont {K.}~\bibnamefont {Gietka}},
  \bibinfo {author} {\bibfnamefont {A.}~\bibnamefont {Golnik}}, \bibinfo
  {author} {\bibfnamefont {P.}~\bibnamefont {Kossacki}}, \bibinfo {author}
  {\bibfnamefont {T.}~\bibnamefont {Jakubczyk}}, \bibinfo {author}
  {\bibfnamefont {W.}~\bibnamefont {Pacuski}}, \bibinfo {author} {\bibfnamefont
  {C.}~\bibnamefont {Kruse}}, \ and\ \bibinfo {author} {\bibfnamefont
  {D.}~\bibnamefont {Hommel}},\ }\href
  {http://przyrbwn.icm.edu.pl/APP/PDF/120/a120z5p15.pdf} {\bibfield  {journal}
  {\bibinfo  {journal} {Acta Phys. Pol. A}\ }\textbf {\bibinfo {volume}
  {120}},\ \bibinfo {pages} {877} (\bibinfo {year} {2011})}\BibitemShut
  {NoStop}%
\end{thebibliography}
%

\end{document}